\documentclass[preprint2]{aastex}

\slugcomment{\it Submitted to the Astronomical Journal}
\shortauthors{Drozdovsky et al.}
\shorttitle{{\it HST}/ACS Grism Parallel Survey II}
 
\newcommand\cf{{cf.,~}}
\newcommand\ie{{i.e.,~}}
\newcommand\eg{{e.g.,~}}
\newcommand\etal{{et al.}}
 
\newcommand\kms{\ifmmode {\rm\,km\,s^{-1}}\else
    ${\rm\,km\,s^{-1}}$\fi}
\def\ergcm2s{\ifmmode {\rm\,ergs\,cm^{-2}\,s^{-1}}\else
    ${\rm\,ergs\,cm^{-2}\,s^{-1}}$\fi}
\def\ergAcm2s{\ifmmode {\rm\,ergs\,cm^{-2}\,s^{-1}\,\AA^{-1}}\else
    ${\rm\,ergs\,cm^{-2}\,s^{-1}\,\AA^{-1}}$\fi}
\newcommand\ergs{\ifmmode {\rm\,ergs\,s^{-1}}\else
    ${\rm\,ergs\,s^{-1}}$\fi}
\newcommand\kmsMpc{\ifmmode {\rm\,km\,s^{-1}\,Mpc^{-1}}\else
    ${\rm\,km\,s^{-1}\,Mpc^{-1}}$\fi}
 
\newcommand\lya{Ly$\alpha$}

\newcommand\oii{[\ion{O}{2}] $\lambda$3727}
\newcommand\oiii{[\ion{O}{3}] $\lambda$5007}

\newcommand\oiiipair{[\ion{O}{3}] $\lambda \lambda$4959,5007}
\newcommand\nii{[\ion{N}{2}] $\lambda \lambda$6548,6584}
\newcommand\sii{[\ion{S}{2}] $\lambda \lambda$6716,6731}
 
\def\spose#1{\hbox to 0pt{#1\hss}}
\newcommand\simlt{\mathrel{\spose{\lower 3pt\hbox{$\mathchar"218$}}
     \raise 2.0pt\hbox{$\mathchar"13C$}}}
\newcommand\simgt{\mathrel{\spose{\lower 3pt\hbox{$\mathchar"218$}}
     \raise 2.0pt\hbox{$\mathchar"13E$}}}
 
\begin{document}

\title{
The {\it HST}/ACS Grism Parallel Survey: II. First Results \\
and a Catalog of Faint Emission-Line Galaxies at $z \leq 1.6$}

\author{Igor Drozdovsky\altaffilmark{1,2},
Lin Yan\altaffilmark{1},
Hsiao-Wen Chen\altaffilmark{3},
Daniel Stern\altaffilmark{4}, \\
Robert Kennicutt, Jr.\altaffilmark{5},
Hyron Spinrad\altaffilmark{6},
\& Steve Dawson\altaffilmark{6}}

\altaffiltext{1}{{\it Spitzer} Science Center, California Institute
of Technology, MS 100-22, Pasadena, CA 91125}

\altaffiltext{2}{Astronomical Institute of St.Petersburg University,
Petrodvoretz, 198504, Russia}

\altaffiltext{3}{Massachusetts Institute of Technology, Cambridge,
MA 02138}

\altaffiltext{4}{Jet Propulsion Laboratory, California Institute of
Technology, Mail Stop 169-506, Pasadena, CA 91109}

\altaffiltext{5}{Steward Observatory, University of Arizona, 933 N.
Cherry Ave., Tucson, AZ 85721}

\altaffiltext{6}{Department of Astronomy, 601 Campbell Hall, University
of California, Berkeley, CA 94720}

\begin{abstract}

We present the first results from the {\it Hubble Space Telescope}
({\it HST}) Advanced Camera for Surveys (ACS) Grism Parallel Survey,
a large program obtaining deep, slitless ACS grism spectroscopy of
high-latitude {\it HST} parallel fields.
We report on 11 high Galactic latitude fields here, each
with grism integration times $>12$~ks.  We identify 601 compact
emission line galaxies at $z \leq 1.6$, reaching emission lines to a flux
limit of $\ga 5 \times 10^{-18}\, \ergcm2s$ ($3\sigma$).  We determine
redshifts by cross correlation of the target spectra with template
spectra, followed by visual inspection.
We measure star formation
rates from the observed \oii, \oiii, and H$\alpha$ line fluxes.  Follow-up
observations with the Keck telescope of one of the survey fields confirms
our classification and redshifts with $\sigma(z)\simeq 0.02$.
This is one of the deepest emission line surveys to date, covering a
total area of 121~arcmin$^2$.
The rough estimate of the co-moving number density of emission-line galaxies
in our survey at $0.3 < z < 1.3$ is
$\sim 4.5 \times 10^{-3}\, h^{-3}_{70}$~Mpc$^{-3}$.
We reach deeper into the emission-line luminosity function than
either the STIS or NICMOS grism parallel
surveys, finding an apparent space density of emission line galaxies
several times higher than those surveys.
Because of the ACS high spatial
resolution, our survey is very sensitive to faint, compact
galaxies with strong emission lines and weak continua.
The ACS grism survey
provides the co-moving star formation density at $z \leq 1.6$ at a high level
of completeness.

\end{abstract}


\section{Introduction}

One of the key unanswered questions of modern cosmology is the
origin and extent of the decline in the star formation rate of the Universe
between $z\sim1.5$ and the present epoch \citep[\eg][]{Gallego:95, Tresse:98,
Lilly:96, Cowie:99, Hippelein:03}.
Based on studies of star-forming and evolving galaxies as a function
of look-back time, a picture has emerged in which more massive galaxies
undergo a larger fraction of their star formation at earlier times than
less massive ones \citep[\eg][]{Cowie:96, Gav_Sco:96, Cimatti:04,
Glazebrook:04, Kodama:04}.
This potentially presents a challenge for
existing models of galaxy formation
\citep[\eg][]{Brinchmann:04}.
To fully address this question, it is necessary
to obtain a comprehensive sample of star-forming galaxies
of different masses and morphological types at a broad range of
redshifts; deep spectroscopic surveys
provide ideal samples for this endeavor.


The low sky background and high spatial resolution afforded by space
makes emission line surveys from above the atmosphere particularly
sensitive and powerful tools for studying galaxy formation and
evolution.  Ground-based, objective-prism programs, such as the Kitt Peak
International Spectroscopic Survey \citep[KISS;][]{Gronwall:04} have been
quite effective at identifying bright, low-redshift H$\alpha$ emitters
over wide areas of sky.  To date, KISS has identified 2266 emission-line
objects over 182 deg$^2$, to a limiting flux of $1 \times 10^{-15}\,
\ergcm2s$.  Ground-based surveys have also been effective at identifying
faint, high-redshift \lya\ emitters in gaps between the telluric night-sky
lines \citep[\eg][]{Hu:98, Rhoads:00, Kodaira:03, Stern:04b}, reaching
typical limiting line fluxes of $\approx 2 \times 10^{-17}\, \ergcm2s$.
Slitless, grism surveys from space provide the opportunity to identify
fainter lines unobstructed by telluric OH emission.  Furthermore,
parallel programs with the {\it Hubble Space Telescope} ({\it HST})
provide, at no cost to the observatory efficiency, surveys which are
less susceptible than pencil beam surveys to
the bias induced by cosmic variance \citep[\eg][]{Cohen:00}.

Previous {\it HST} parallel slitless spectroscopic programs with the
Space Telescope Imaging Spectrometer \citep[STIS;][]{Gardner:98,
Teplitz:03} and the Near-Infrared Camera and Multi-Object Spectrograph
\citep[NICMOS;][]{McCarthy:99, Yan:99} have left a valuable scientific
legacy studying faint emission line galaxies out to high redshift.
In particular, the lack of telluric
emission lines allowed the NICMOS Parallel Survey to identify an
impressive census of high-redshift galaxies whose optical features are
shifted into the near-infrared.  ACS, though observing at the same
wavelength regime as STIS,
provides higher spatial resolution data and is a factor of a few more
sensitive to emission lines,
thus providing a significant increase in our ability to study 
faint emission-line galaxies \citep{Pirzkal:04}.

In this paper we present the analysis of eleven deep spectroscopic fields
from the ACS parallel survey.  ACS slitless spectra provide
unprecedented sensitivity in the range $5500 < \lambda < 10500$
\AA, where ground-based spectroscopy is challenged by the night sky.
We discuss a sample of $z \leq 1.6$ galaxies selected by the presence
of \oii, \oiii\ and/or H$\alpha$ emission features.  Our survey is
well-suited for exploring the faint end of the star-forming galaxy
luminosity function,
identifying \lya\ emitting galaxies at $4 \simlt z \simlt 7$,
and for studying the spatially-resolved star formation rate in individual
galaxies.  This paper presents the first results from our survey.
In \S2 we describe our observation and data reduction methodologies.
In \S3 we describe how redshifts are determined and present a comparison
of our ACS grism redshifts to ground-based observations obtained at Keck
Observatory.  In \S4 we present initial results from our survey, including
measurements of star formation rates for individual galaxies (\S4.2) and
ACS morphologies of actively star-forming galaxies (\S4.3).  Our results
are summarized in \S5.  The first paper in this series describes our
data reduction scheme in detail \citep[Paper~I;][]{Chen:04}; future
papers will address the faint end of the luminosity function for star-forming
galaxies (Paper~III; Drozdovsky \etal, in preparation), and the early-type
galaxy sample at $ 0.6 \simlt z \simlt 1.3  $ detected
with the grism spectra (Paper~IV; Yan \etal, in preparation).

Throughout this paper, unless otherwise specified, magnitudes refer to the
Vega system and we adopt a flat, $\Lambda$-dominated universe (H$_0=70\,
h_{70}$~km s$^{-1}$ Mpc$^{-1}$, $\Omega_M=0.3$, $\Omega_{\Lambda}=0.7$).

\section{{\it HST}/ACS Observations}

All {\it HST} data presented here were obtained with the Wide Field Camera
(WFC) on ACS.
The observing program was designed to acquire a pair of images at each
pointing:  a direct image taken with a broad-band filter (typically F775W
or F814W) and a dispersed image taken with the G800L grism
(see Table~\ref{t:Fields}).  The direct
images are important both for registering the grism frames to a common
origin and for the 0$^{\rm th}$ order wavelength calibration of the grism
spectra.  The WFC/G800L grism has a mean dispersion of $\sim 40$~\AA\,
pix$^{-1}$ in the first order \citep{ACS_DOC}. The actual, observed resolution is a
function of the apparent image size convolved with the WFC point spread
function (PSF).  From the observations described herein, the realized
spectral resolution, $R_{\lambda}$, is $\sim 80$ to $\sim 150$.
Small variations in the PSF
due to changes in the optical telescope assembly (\cf {\em breathing})
and longer term changes in the internal structure of ACS introduce small
variations in the maximal achievable resolution.

\subsection {Parallel observing mode and scheduling}

Approximately half of the observations for this investigation were obtained
with {\it HST} operating in parallel mode as part of GO/PAR program
9468\footnote{
Information about the observations can be gleaned directly from the
STScI WWW pages linked to the program ID.} (P.I. Lin Yan).
While this allowed us to collect far more data than would be possible
in a single primary program, it limited our ability to plan and execute
the observations in a manner that optimized the scientific return.
Since July 2002, the ACS/WFC parallels were scheduled during much
of the time for which NICMOS or STIS were the primary instruments.
The observing algorithm for this program was quite simple. In each full
orbit one of the following exposure sequences was selected: F606W and/or
F814W imaging followed by G800L grism observations with exposure times
of approximately 500~s.  The preferred ratio of exposure times
was F814W:F606W:G800L = 1:1:3.  The selection of the exposure sequence
for any given orbit was nearly random, but was weighted in favor of the
spectroscopy.  The dither step and orientation depended
upon the primary observing programs, and we preferentially use programs
with fixed orientation and small dither offsets.

Another half of our data were obtained from the {\it HST} Archive
as part of guest observer parallel GO/PAR program 9482 (P.I. James
Rhoads), and data for the J~08:08+06:43 field are from GO program 9405
(P.I. Andy Fruchter).  The basic observing approaches were similar to
ours, though with alternate filter selections (see Table~\ref{t:Fields}).

During the period from July 2002 to October 2003, approximately 800 ACS
grism exposures (roughly 200~orbits) were observed.  These were distributed
in $\sim 40$ independent pointings with integration time of each individual
grism exposure ranging from 300~s to 1200~s.  Rather than adopt a strict latitude
cutoff we chose to reduce all of the data and reject those with high
stellar densities.  Depth of the grism images is the major limitation of
the study.  For this study we choose 11 fields with total grism exposure
time more than $>12$~ks, covering about 121 square arcminutes.

\begin{figure}[tb]
\vbox{\includegraphics{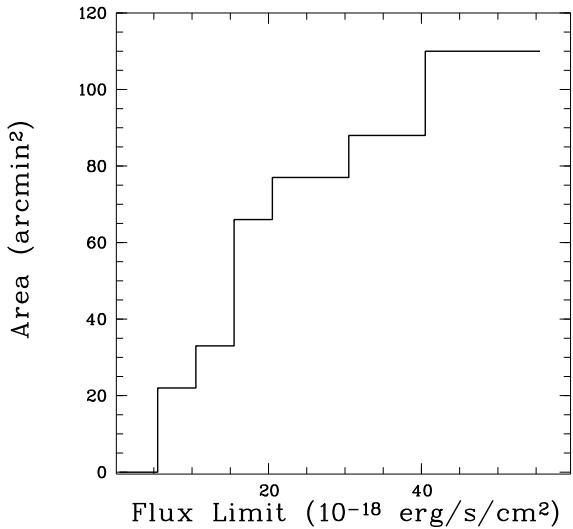}}
\vspace{6.9cm}
\caption{Area-depth histogram for the ten ACS grism parallel fields
discussed here.  Flux limits refer to 3$\sigma$ limits in 4 pix apertures;
this aperture reasonably represents the ACS/WFC PSF.
}

\label{f:Fldepth}
\end{figure}


The final depth achieved varied from field to field.  We define our
limiting depths as $3\sigma$ within a 4 pixel aperture.  This aperture
reasonably represents the area of the ACS/WFC PSF.  Fig.~\ref{f:Fldepth}
shows an area-depth histogram for our survey.  Our median limiting
depth is $\sim 1.6 \times 10^{-17}\, \ergcm2s$  and our deepest
three fields reach a depth of $\sim 5 \times 10^{-18}\, \ergcm2s$, which
is approximately a factor of two deeper than typical
ground-based narrow-band/spectroscopic surveys and $2 - 5$ times deeper than
the HST/NICMOS \citep{McCarthy:99, Yan:99} and STIS \citep{Teplitz:03}
parallel surveys.

\subsection{Data processing and analysis}

The extraction of the spectra from the grism images is a multi-stage
process.  The main steps of the reduction procedure are:  standard {\tt
CALACS} pipeline reduction (\eg bias subtraction); cosmic
ray detection and global sky background removal; combining the 2D images;
generating object catalogs from the direct images; extraction and
calibration of 1D spectra from the coadded grism images; combination
of the 1D spectra for objects observed at different positions and/or
with different orientations. The software used for the spectral extraction
is written by one of us \citep[][Paper~I]{Chen:04}. Paper~I
presents the design and performance of our data reduction software.
We detail our reduction steps below.

\subsubsection{{\tt CALACS} pipeline and image combination}\label{ss:ima_comb}


We begin with the output from the STScI data pipeline, {\tt CALACS}, which
does basic processing of the individual 2D frames.  All data have first
order bias subtraction, dark subtraction, and bad pixel masking applied.
Direct images are then corrected for the flat-field response.
Since the flat-field response is a function of wavelength for each
pixel and so depends on the location of sources in the dispersed data,
the grism data are not flat-fielded.\footnote{
The {\tt aXe} ACS grism reduction code of
\citet{Pirzkal:02} uses a data cube constructed from observations through
narrow-band filters, interpolated to the wavelengths of extracted pixels,
to flat-field grism spectra.  This technique reduces pixel-to-pixel scatter
within
the $6000-8500$~\AA\, wavelength regime from 3\% to 1\%, but shows no
improvement beyond 8500~\AA\, due to a lack of long-wavelength ACS
narrow-band filters.  Since the G800L grism mode remains sensitive to
$\approx 10\, 000$~\AA, we opt to omit flat-fielding during spectral
extraction (see Paper~I for details).}


We combine the {\em direct} images using a modified version of the {\tt
Multidrizzle} package \citep{Koekemoer:02}. {\tt Multidrizzle} is an
implementation
of the  {\tt Blot/Drizzle} technique \citep{Fruchter:02} which provides
an automated method for distortion-correcting and combining dithered
images.  {\tt Multidrizzle} corrects for gain and bias offsets between
WFC chips and identifies and removes cosmic rays and cosmetic defects.
The quality of the image combination relies on accurately determining
the offsets between images.  The original program aligns images using
the World Coordinate System (WCS) header keywords.  Unfortunately,
there are times when those WCS values are inaccurate, resulting in a
misalignment of the final output drizzled product.  We have therefore
enhanced {\tt Multidrizzle} with
the ability to internally verify and update WCS-calculated offsets using
image cross-correlation or point-source matching algorithms.  The derived
shifts and rotations are determined for the geometrically corrected frames
resampled onto a common WCS frame.  The absolute astrometry of the final
drizzled images were verified using guide stars as well
as extragalactic sources from NED. The astrometric uncertainties are
dominated by the accuracy of the coordinates of the guide stars located
in a field area and differ from field to field with a median value of
$\sim 0\farcs2$.

For the {\em grism} data, our ability to stack raw ACS frames is limited by the
substantial geometric distortion and the extent of the offsets between
individual exposures.  According to the ACS Instrument Handbook, the
plate scale changes by 8\% between the two diagonal corners of the field.
This discrepancy is less than a pixel for dither offsets smaller than
10 pixels.  Therefore, we register distorted ACS grism images that have
pointing offsets less than 10 pixels into {\em substacks}, taking care
to mask bad pixels and applying a $5\sigma$ clipping algorithm to remove
hot pixels and residual cosmic rays.  For fields with larger offsets or
varying orientations between exposures, we create multiple substacks.
Spectral extraction is usually performed on these substacks, then
the final coadd is done for 1-dimensional spectra.

Our spectral extraction software (Paper I)
requires a pair of aligned direct
and dispersed images in the original post-pipeline format; \ie each
WFC chip image must reside in a separate FITS files with the original
ACS distortion.  We make use of the {\tt Blot} program, which transforms
direct images back to the reference frame of G800L image stack(s).
We extract
1D spectra from the substacks. In cases with multiple substacks, the 1D
extracted spectra are resampled onto a common grid and weight-averaged
to create the final deep spectrum.


\subsubsection {The Catalog}

We generate object catalogs from the stacked direct image using {\tt
SExtractor} \citep{Bertin:96}, applying a $1.5 \sigma$ per pixel detection
threshold and $8$ pixel minimum area requirement.  Table \ref{t:Phot}
presents an example of the catalog produced from our imaging dataset;
{\em the full catalog of emission line galaxies is produced in the
on-line version of this paper}.
Below we detail some of the listed parameters:

\begin{itemize}

\item {\em Equatorial coordinates} are the barycenter position of a
source, that were measured in a WCS-corrected direct image (see
\S~\ref{ss:ima_comb}).

\item {\em Magnitudes} are measured in Kron-like AUTO magnitudes and
transformed to the Vega photometric system using following zero-points:
26.398 for F606W, 25.256 for F775W, 25.501 for F814W, and 24.326
for F850LP \citep{ACS_DOC}.

\item{{\em a,b} are the major and minor axis lengths of the fitted ellipse
(in units of the $0\farcs05$ ACS/WFC pixels).}

\item{$\theta$ is the position angle of the major axis with respect to
the dispersion direction.}

\item{{\em Concentration}, the \citet{Abraham:96} concentration parameter,
is the ratio between the flux in the central 30\% of the pixels as
compared to the total flux.}

\item{{\em Asymmetry}, the \citet{Abraham:96} point-asymmetry index,
is the absolute sum of the differences between point symmetric pixels
divided by the total flux.}

\end{itemize}

\subsubsection {1-D spectral extraction and calibration }

The extraction and calibration of the 1D spectra was carried out
with custom software developed by Hsiao-Wen Chen and described in detail
in Paper~I.  We briefly describe the methodology here; the
interested reader is directed to Paper~I for details.

We first align sources in the direct image (position $x_i,y_i$) with
the corresponding position of their zeroth orders in the dispersed image
($x_i^0,y_i^0$), adopting the empirically-derived transformation:
\begin{equation}
\begin{array}{ll}
x^0_s = -122.1591+1.00442\cdot x_i - 0.00395\cdot y_i;
\\
y^0_s = 3.6392-0.00014\cdot x_i + 1.00088\cdot y_i
\end{array}
\end{equation}
\noindent
for WFC Chip 1, and
\begin{equation}
\begin{array}{ll}
x^0_s = -115.2942+1.00444\cdot x_i - 0.00352\cdot y_i;
\\
y^0_s = 2.5025-0.00028\cdot x_i + 1.00068\cdot y_i
\end{array}
\end{equation}
for WFC Chip 2.  The extraction region of the spectrum was then computed
using transformations derived as part of the calibration program.
Using the object brightness distribution inside its segmentation map on
the direct image, the corresponding 2D spectra were modeled and overlapping
spectra were de-blended via iterative, multiple-profile fitting procedure
for all spectra in a frame simultaneously.  For objects close to the
edges of the dispersed images, only partial spectra were extracted.
For faint sources, only first order light is considered; for bright
sources, we improve the signal by including higher orders.  The rms
accuracy of the wavelength calibration for G800L is approximately $7
$\AA\/ \citep{Pasquali:02}.
We flux calibrate the extracted spectra using
calibration curves derived by \citet{Pirzkal:02} from observations of
white dwarfs and Wolf-Rayet stars.  The accuracy of the spectrophotometry
is limited by uncertainties in the wavelength calibration, the various
detector flat-field effects, the object deblending, and
variations in the quantum efficiency within individual
pixels. We estimate that the absolute flux calibration is accurate to
approximately 5\% from 5000~\AA\ to 9000~\AA.

Spectra were also extracted in a parallel effort using the {\tt
aXe} software developed at ST-ECF \citep{Pirzkal:02}.  The {\tt aXe}
software follows a similar strategy to our code, except it does not 
perform any deblending due to higher orders from bright objects.
Spectra are extracted in weighted boxes, with
flat-fielding performed based on observations through narrow-band filters,
interpolated to the wavelengths of the extracted pixels.  Flux calibration
uses the same calibration curves as used by us.  The overall results of
the extraction and analysis using the {\tt aXe} and our software were
similar, except in cases of overlapping spectra where the de-blending
included in our code produces significantly cleaner extractions.

\section{Redshift Identifications}

\begin{figure*}
\centering{
\vbox{\includegraphics{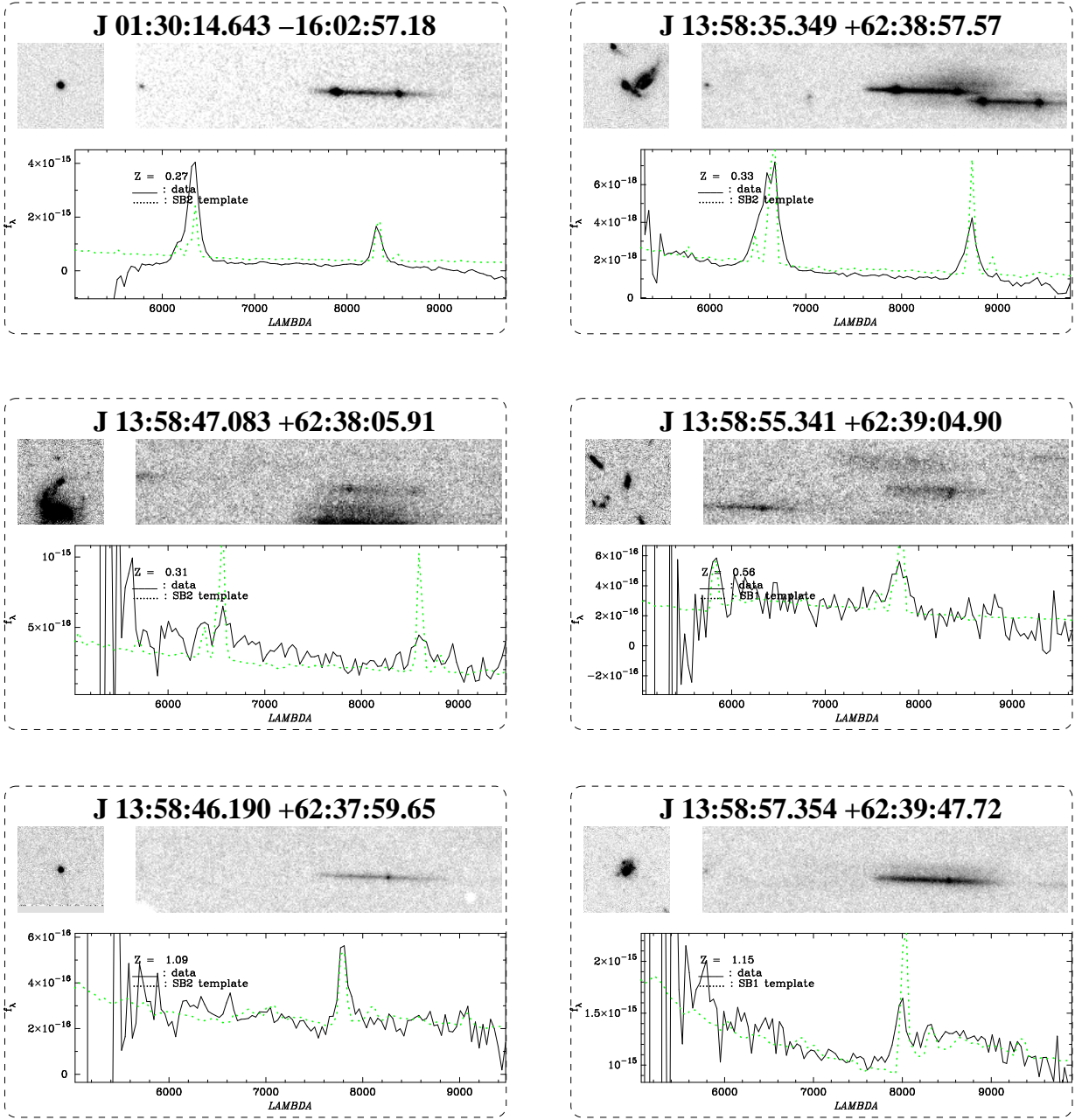}}\par
\vspace*{16.0cm}
}
                                                                                
\caption{Typical emission-line galaxies detected from the ACS Grism
Parallel Survey, showing direct images, dispersed images, and extracted
1D spectra. All spectra are shown over observed wavelength range in
flux units of erg~s$^{-1}$~cm$^{-2}$~\AA$^{-1}$.
The dotted lines indicate the best-fitting spectral template.
Data are sorted by the reliability of their line identifications: the
top row shows example of robust, class `a' redshift identifications, the
middle row shows examples of likely, class `b' redshift identifications,
and the bottom row shows less secure, class `c' redshift identifications.
The postage stamp images are 5\arcsec\ on a side, oriented as per the
original data.}

\label{f:elgs}
\end{figure*}

We next cross-correlate the extracted, 1D spectra with a set of stellar
and galactic spectral templates in order to deduce an approximate spectral
type and redshift for each source.  Sources were divided into four groups:
Galactic stars, emission-line galaxies (ELGs), early-type galaxies,
and galaxies with a \lya\ break at $4<z<7$.  In this paper we present
analysis of only the ELGs.  Emission-line fluxes and equivalent widths
were measured using Gaussian fits to the line and polynomial fitting of
the underlying continuum, performed using the ESO-MIDAS {\tt Alice} package.
Fig.~\ref{f:elgs} presents direct images and spectra of a sample of 6
ELGs from this survey.  Galaxies were selected to illustrate the range
of data quality.  Tables \ref{t:Phot} and \ref{t:Spec} contain the catalog
of imaging and spectroscopic properties of the ELGs detected in
our fields\footnote{The complete version of these tables are in the electronic
edition of the Journal. The printed edition contains only a sample.}.

\subsection {The identification of emission lines}

Although $\sim80\%$ of objects with visible emission lines were
successfully classified with the
cross correlation technique, the accuracy of the redshift determination
was low and required visual verification. We compared the ACS
spectra to star-forming galaxy templates from \citet{Kinney:96} smoothed
to the grism resolution. To convince ourselves that any detected emission
line was real, we eliminated all possible false signals (\eg zero order
images, persistent images, cosmic rays, etc.).  For final spectral
classification, three criteria were used:  identified emission line(s),
morphology of the emission line(s), and the continuum spectral energy
distribution (SED).

To be specific, three major emission features could be present
at our observed wavelength range:  H$\alpha$ blended with \nii\
(for galaxies at $z<0.5$),
the unresolved \oiiipair\ doublet\ + H$\beta$
($0.5<z<1.0$), and \oii\ ($0.6<z<1.7$)\footnote{Galaxies with \lya\ emission
at $z>4$ are not considered here.}.  Many sources show multiple emission
lines, allowing unambiguous redshift identifications.  We note that
since the ratio of \oiiipair\ flux to both H$\alpha$ and \oii\ flux
varies significantly in both the local universe \citep{Terlevich:91,
Kennicutt:92, Izotov:94} and at high redshift \citep[\eg][]{Kobulnicky:99,
Teplitz:00, Pettini:01}, line flux ratios do not provide a strong
redshift indicator.

Galaxies with only a single emission line require more attention.
We assume that isolated emission lines are either H$\alpha$ or \oii.
Since the [\ion{O}{2}]:[\ion{O}{3}] ratio can vary from 0.1 to
10, and the [\ion{O}{3}]:H$\alpha$ ratio can vary from 0.33 to 1
\citep{Kennicutt:92}, the non-detection of [\ion{O}{3}] in an H$\alpha$-
or [\ion{O}{2}]-emitting galaxy is possible.  We choose between H$\alpha$
and [\ion{O}{2}] based on the relative position and strength of the line,
the continuum SED,
as well as the brightness and morphology of the host galaxy in the direct
image.  For example, a single emission line at wavelengths shortward of
7800~\AA\ that would imply either an H$\alpha$-emitter at a low redshift
($z<0.2$) or an [\ion{O}{2}] emitter at moderate redshift ($z \simlt 1$).
An H$\alpha$-emitter might be expected to appear noticeably larger
and brighter in the direct image, possibly showing resolved structure.
Since the space density of low-redshift, star-forming compact dwarf
galaxies is low, a single, red emission line in a faint galaxy
(with a blue rising continuum at shorter wavelength) is most
likely [\ion{O}{2}] unless there is evidence to the contrary.

Line morphology can also provide a useful redshift diagnostic.  At our
spectral resolution, the [\ion{O}{3}] line is generally asymmetric due
to blending with H$\beta$, allowing us to secure line identifications
based on the line morphology. On the other hand, \nii\ are not resolved
from H$\alpha$ at our resolution, particularly at the low redshifts at
which these lines are observed. Some of the compact
low-redshift sources, however, reveal the noticeable
\sii\ doublet.

We assign a quality (reliability) flag to all redshift estimations
based on the number of detected emission lines and significance of
their identifications.  Quality flag `a' indicates that there are two
or more emission lines identified and that the [\ion{O}{3}]+H$\beta$
blend shows a clear asymmetry; this flag indicates that the redshift
is secure.  A value of `b' is assigned to galaxies for which there is a
strong reason for the assignment.  Specifically, such a quality assignment
implies we observe multiple emission lines but doubts remain as to their
identifications.  Quality flag `c' indicates even greater uncertainty
in the redshift identification, typically indicating that only one line
has been detected.  A non-detection of the second line can be due to
several reasons:  the most common circumstance is that the galaxy is
either at $z<0.2$ or at $z>0.8$, and [\ion{O}{3}] is in a region of poor
spectroscopic sensitivity ($\lambda<6000$~\AA\ or $\lambda>9000$~\AA).
In a few cases, the spectral range is truncated since the galaxy lies
near the edge of the field of view.

\begin{figure}[tb]
\vbox{\includegraphics{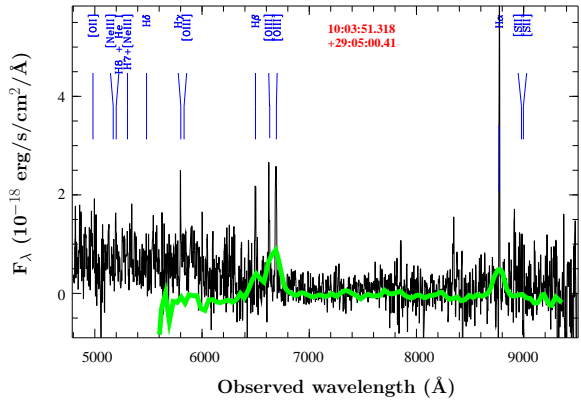}}
\vspace{5.3cm}

\caption{Keck/LRIS spectrum (thin line) of one of the ELGs discovered in this
survey, overlaid on the ACS grism discovery spectrum (thick grey line).
Both LRIS and ACS spectra produced similar redshift determinations, within the
expected range considering the $\sim 25$ times lower spectral resolution
of the ACS grism data.}

\label{f:Keck}
\end{figure}

\subsection {Comparison with Keck spectroscopy}\label{s:lris}

To test the accuracy of the redshifts and study potentially-interesting
faint sources, on UT 2004 March 19 we obtained spectroscopy of two
slitmasks targeting ACS-selected galaxies (field J~10:03$+$29:06)
with the Low Resolution Imager
and Spectrograph (LRIS; Oke et al. 1995) on the Keck~I telescope.
These observations were obtained in non-photometric conditions and
integration times totaled one hour per slitmask, split into three
dithered 1200~s exposures.  LRIS is a dual-beam spectrograph:  we used
the D680 beam-splitter, the 300 lines mm$^{-1}$ grism ($\lambda_{\rm
blaze} = 5000$~\AA; $\Delta \lambda_{\rm FWHM} = 9.0$~\AA) on the
blue arm, and the 400 lines mm$^{-1}$ grating ($\lambda_{\rm blaze}
= 8500$~\AA; $\Delta \lambda_{\rm FWHM} = 6.4$~\AA) on the red arm.
Data were processed using standard techniques.  Because the night was
not photometric, we used archival sensitivity functions dating from
March 2002 for relative flux calibration.

Unfortunately due to poor weather conditions, we were unable to study
the more extreme, faint sources identified from ACS. However, the LRIS data
proved quite useful for verifying the accuracy of the ACS-derived
redshifts.  A total of 11 ACS-selected ELGs were observed,
out of which one was of quality `a', five were of quality `b', and
five were of quality `c'.  In all cases, the ACS and Keck redshifts
were consistent.  The mean redshift difference is negligible, $\langle
z_{\rm ACS} - z_{\rm Keck} \rangle = -0.01 \pm 0.02$.  Fig.~\ref{f:Keck}
presents both the ACS and the Keck spectra of
one of the sources, J\,10:03:51.318$+$29:05:00.41, at redshift $z=0.337$.
Note that the H$\beta$/[\ion{O}{3}] complex is barely resolved by ACS,
but is well-resolved by Keck.

Artificial object tests and comparison of our redshift estimations with
ones measured using the higher spectral resolution Keck/LRIS spectra
indicate that manual inspection correctly finds the redshifts for 
over 95\% of the spectra.

\begin{figure}[tb]
\vbox{\includegraphics{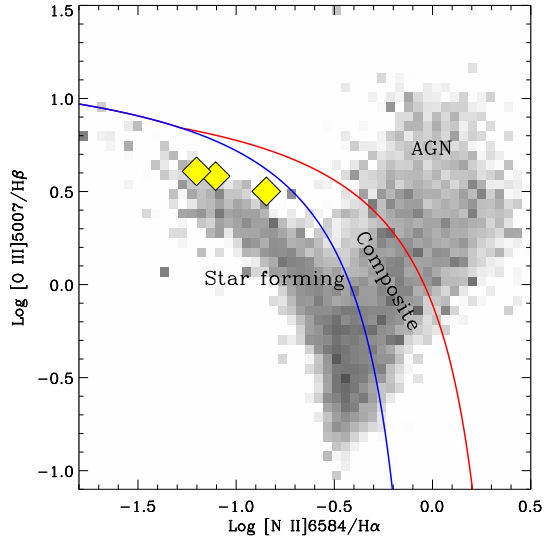}}
\vspace{7.0cm}

\caption{The location of three ELGs (diamonds) based on the Keck/LRIS
spectra in the BPT line-ratio diagram from \citet{Brinchmann:04}.
Two lines shows the empiric classification of the ELGs by activity
type.
}\label{f:BPT}
\end{figure}

The Keck/LRIS spectra confirms the high efficiency of the
ACS grism selection method, which yields a high fraction of strong-lined
galaxies. Namely, 7
out of 11 observed ELGs (64\%) have EW(\oiii)$>100$~\AA\,.
Such galaxies tend to be either star-burst galaxies with little extinction,
or AGN with high-excitation spectra.
The general way to classify the ELGs by activity type
(AGN versus starburst) is based on the
emission line ratios \citep[][hereafter BPT]{Baldwin:81}.
We classify three of these strong-lined sources as
star-forming galaxies based on their \oiii/H$\beta$ versus
\ion{N}{2}\,$\lambda6584/$H$\alpha$ line-ratio,
shown in Fig.~\ref{f:BPT}.
For the remaining galaxies, comparison of their \oiii/H$\beta$
ratio with their \oii/\oiii\, ratio, immeasurably weak
[\ion{Ne}{3}]\,$\lambda3826$ line, and narrow emission lines
confirm that they are also starburst systems.

\begin{figure}[tb]
\vbox{\includegraphics{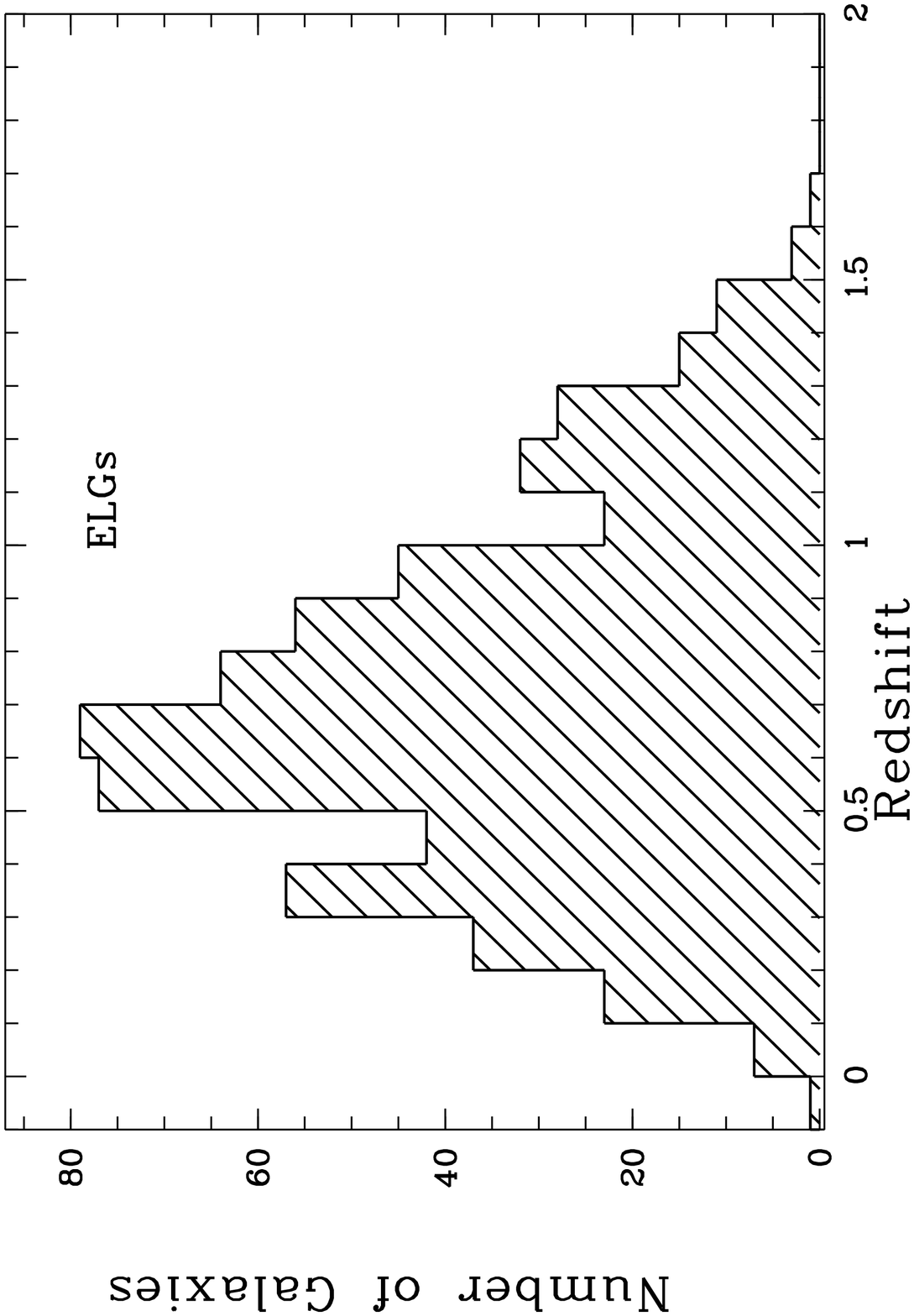}
      \includegraphics{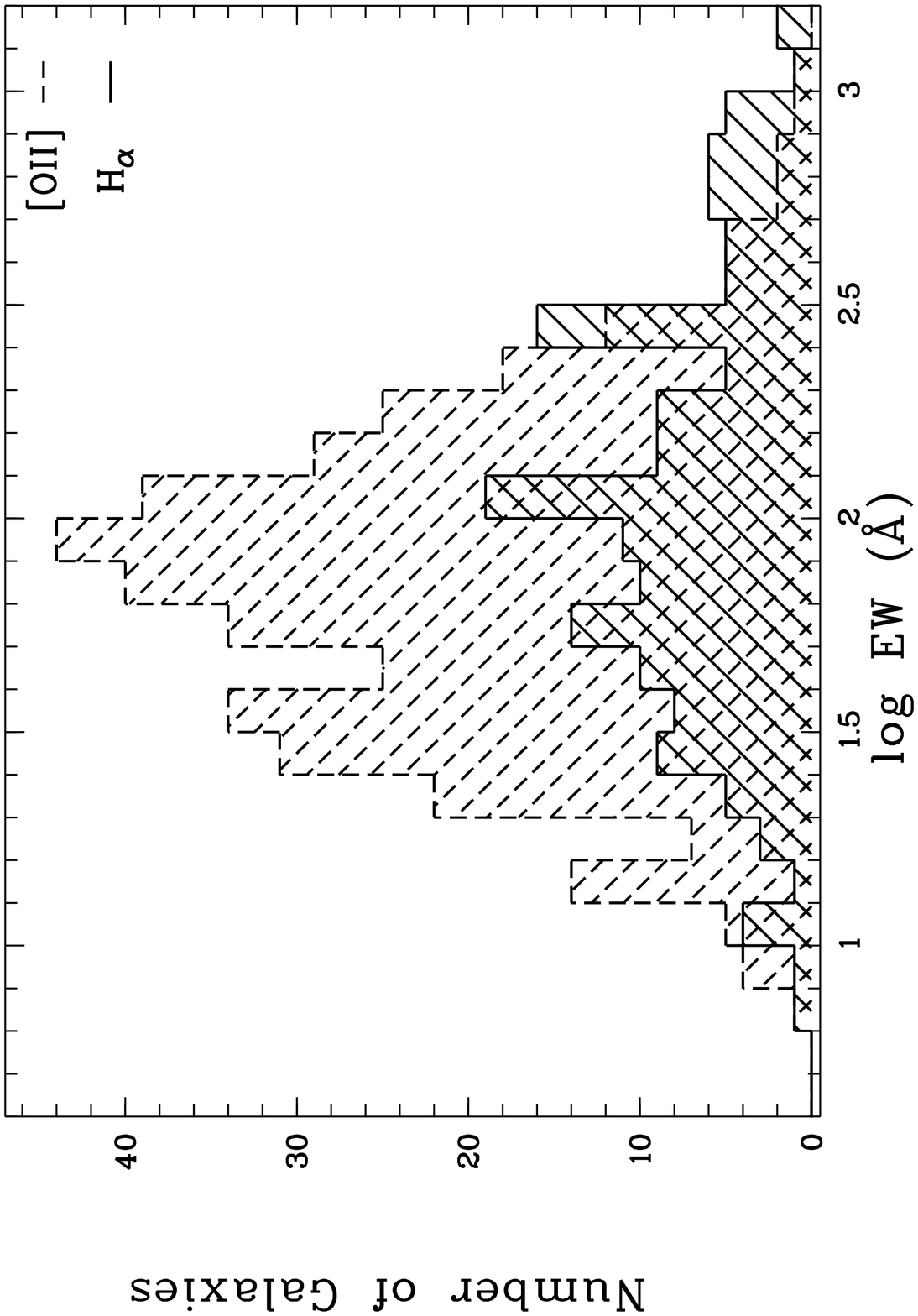}
}
\vspace{10.8cm}

\caption{Redshift distribution (top) and observed equivalent width
distribution (bottom) for detected ELGs.}

\label{f:Zhist}
\end{figure}

\begin{figure}[tb]
\vbox{\includegraphics{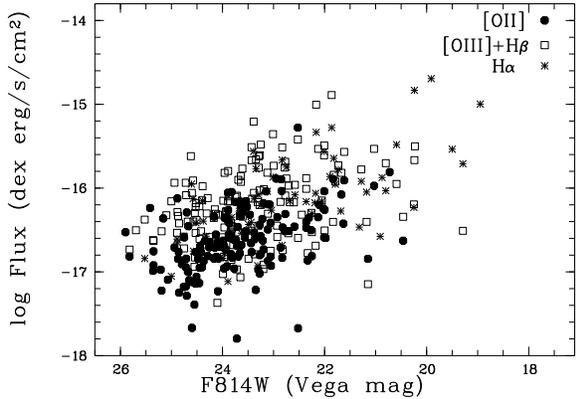}}
\vspace{5.2cm}

\caption{Emission line fluxes versus F814W magnitude.  The symbols
indicate different emission features:  asterisks indicate H$\alpha$
emission, empty squares indicate [\ion{O}{3}] emission, and filled
circles indicate [\ion{O}{2}] emission.  The ACS grism survey identifies
galaxies to very faint continuum brightness levels.}

\label{f:elfl}
\end{figure}

\begin{figure}[bt]
\vbox{\includegraphics{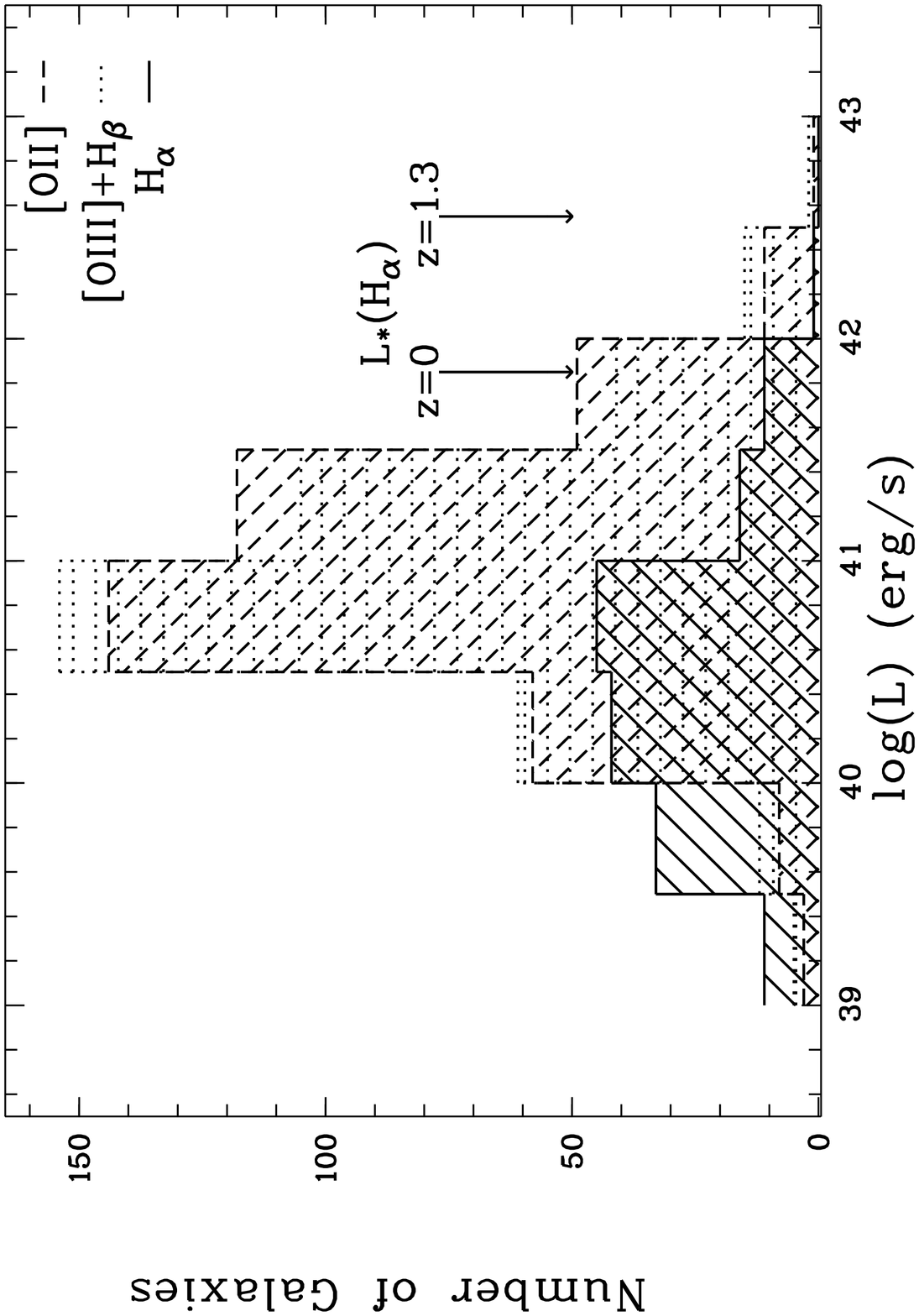}}
\vspace{5.2cm}

\caption{Histogram of H$\alpha$ (solid line), [\ion{O}{3}]+H$\beta$
(dotted line) and [\ion{O}{2}] (dashed line) luminosities in the ACS Grism
Parallel Survey.   Vertical lines indicate the characteristic H$\alpha$
line luminosity for the local universe (Gallego \etal 1995) and at $z =
1.3$ (Yan \etal 1999).}

\label{f:Lhist}
\end{figure}

\section{Results}

\subsection {Areal coverage and depth }

Based on the analysis of eleven fields, we detect 601 galaxies revealing
significant emission features, corresponding to a surface density of
$\sim 5$ ELGs arcmin$^{-2}$.
Among them we identify 166 galaxies with
H$\alpha$ emission, 406 galaxies with [\ion{O}{3}]+H$\beta$ emission, and 401
galaxies with [\ion{O}{2}] emission.  Our survey is most sensitive to
emission line sources with angular sizes (in the dispersion direction)
of less than 1 arcsec and with broad-band magnitudes of F814W $\simlt
26$ mag.  Fig.~\ref{f:Zhist} shows the redshift and equivalent width
distribution for the survey and Fig.~\ref{f:elfl} plots line flux fluxes
against F814W magnitude.  Our ELG sample has a median redshift of
0.66.  At low redshifts, the fall off in H$\alpha$-emitting galaxies is
attributed to the small volume covered by survey.  At high redshifts,
the number of sources falls as the sensitivity drops and our primary
rest-frame optical features enter the near-infrared.
There is also some plunge in the number of detected single-line galaxies
at $0.4<z<0.5$ and $1.0<z<1.1$,
when second H$\alpha$ and [\ion{O}{3}] lines move into the
near-IR.

\begin{figure}[tb]
\vbox{\includegraphics{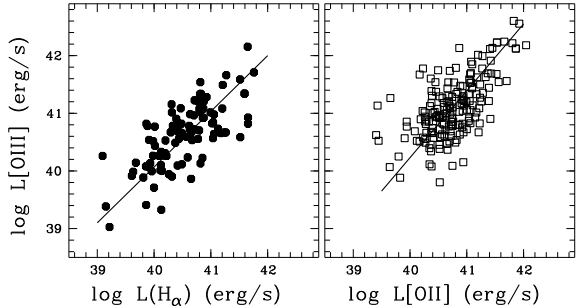}}
\vspace{4.2cm}

\caption{[\ion{O}{3}]+H$\beta$ line-blend luminosities plotted against
H$\alpha$ and [\ion{O}{2}] luminosities for the ACS Grism Parallel Survey.
Lines show the best fit correlation derived from orthogonal regression.
While the scatter is considerable, a correlation is obvious.}

\label{f:O3vsHaO2}
\end{figure}

\subsection {The emission-line luminosity}

The distribution of the emission-line luminosities, ignoring
dust extinction corrections, is presented in Fig.~\ref{f:Lhist}.
The median H$\alpha$ line luminosity of galaxies in our survey is
$2.7\times10^{40}\, h_{70}^{-2}$~ergs s$^{-1}$; this is 26 times fainter
than characteristic luminosity L$^{*}_{{\rm H}\alpha}=7.1\times10^{41}\,
h_{70}^{-2}$~ergs s$^{-1}$ derived from surveys of the local universe
\citep[\eg][]{Gallego:95, Tresse:98}, and 132 times fainter than
L$^{*}_{{\rm H}\alpha}=3.6\times10^{42}\, h_{70}^{-2}$~ergs s$^{-1}$ at
$z=1.3$ \citep{Yan:99}.  The median [\ion{O}{2}] line luminosity is
$8.5\times10^{40}\, h_{70}^{-2}$~ergs s$^{-1}$.  The median [\ion{O}{2}]
luminosity for the $z>0.6$ sample is two times fainter than the local
L$^{*}_{[{\rm OII}]}=1.7\times10^{41}\, h_{70}^{-2}$~ergs s$^{-1}$ derived by
the Universidad Complutense de Madrid survey
\citep[UCM;][]{Gallego:02}.
The median [\ion{O}{3}]+H$\beta$ line-blend luminosity can be compared
with single \oiii\, line luminosity
L$^{*}_{[{\rm OIII}]5007}=1\times10^{42}\, h_{70}^{-2}$~ergs s$^{-1}$
calculated by \citet{Hippelein:03} for galaxies in the
redshift range $0.62<z<0.65$. Our cumulative [\ion{O}{3}]+H$\beta$
value of
L$^{*}_{[{\rm OIII}]+{\rm H}\beta}=7.4\times10^{40}\, h_{70}^{-2}$~ergs s$^{-1}$
is again 14 times fainter.
Our survey
is clearly reaching down to the faint end of the emission-line luminosity
functions.
Our sample selects star-forming galaxies over a wide range of luminosity,
from faint emission-line galaxies at low redshifts to luminous L$^{*}$
galaxies at high redshifts.



\subsection {The emission-line luminosity and star formation rate}

H$\alpha$ emission is a classic indicator of star formation because it
traces the ionizing flux from hot stars.  Assuming Case B recombination
and a Salpeter initial mass function over the mass range $0.1 < M /
M_\odot < 100$, we adopt the calibrations of \citet{Kennicutt:98}
between star formation rate (SFR) and H$\alpha$ luminosity:
\begin{equation}
{\rm SFR}_{\rm H\alpha}\, ({\cal M}_{\odot}\,{\rm yr}^{-1}) =
   7.9\times10^{-42}\cdot L_{\rm H\alpha}\, {\rm (erg\,s^{-1}).}
\end{equation}
Assuming the average H$\alpha$ to [\ion{O}{2}] flux ratio of 0.45,
the \citet{Kennicutt:98} relation implies
\begin{equation}
{\rm SFR}_{\rm [OII]}\, ({\cal M}_{\odot}\,{\rm yr}^{-1}) =
   1.4\times10^{-41}\cdot L_{\rm [OII]}\, {\rm (erg\,s^{-1}),}
\end{equation}
subject to the considerable scatter in this flux ratio found in surveys
of local galaxies, likely associated with variations in the metallicity
and star formation histories of individual galaxies.  Furthermore, we
note that since no extinction corrections have been applied, the derived
luminosities and star formation rates should be considered lower limits.


While H$\beta$ line is also a good tracer of star formation in galaxies
\citep[\eg][]{Kennicutt:83}, it is blended with the \oiiipair\ doublet
in our low-resolution spectra.  The strength of the oxygen lines,
however, also correlate with the formation rate of massive stars,
and may be used as a proxy for the SFR in cases where [\ion{O}{2}]
and H$\alpha$ are unavailable \citep[\eg][]{Teplitz:00, Hippelein:03}.
Due to its high ionization level, the luminosity of [\ion{O}{3}]
depends strongly upon the temperature of the ionized gas, which in
turn depends upon the metallicity of the galaxy.  In order to convert
the [\ion{O}{3}]$+$H$\beta$ line blend into a star formation rate, an
averaged intensity ratio between these lines and H$\alpha$ must therefore
be established.  Having detected and measured a large number of ELGs,
revealing either the [\ion{O}{3}] and H$\alpha$ or the [\ion{O}{3}] and
[\ion{O}{2}] line pairs, we can test the correlation between [\ion{O}{3}]
luminosity and SFR with the understanding that the analysis is inherently
limited by the unknown properties of the galaxies.

Fig.~\ref{f:O3vsHaO2} compares the [\ion{O}{2}], [\ion{O}{3}], and
H$\alpha$ luminosities.  Assuming the average H$\alpha$:[\ion{O}{2}]
ratio of $\sim0.6$, the median [\ion{O}{3}]:H$\alpha$ ratio of $\approx
1.3$ we derive from 127 ACS grism ELGs at $z \simlt 0.6$ is in agreement
with the median [\ion{O}{3}]:[\ion{O}{2}] $\approx 2.2$ ratio we derive
from 245 ACS grism ELGs at $0.5 \simlt z \simlt 1.0$.  Using these ratios
and equations (3) and (4), we derive
\begin{equation}
{\rm SFR}_{\rm [OIII]+H\beta}\, ({\cal M}_{\odot}\,{\rm yr}^{-1}) \simeq
   6\times10^{-42}\cdot L_{\rm [OIII]+H\beta}\, {\rm (erg\,s^{-1})}.
\end{equation}
We expect this SFR indicator to have the largest scatter of the three
considered; consequently, we only use it for the 47 galaxies in our
sample (less than 8\%) which reveal only a single [\ion{O}{3}] line and lack
other emission lines.

\begin{figure}[tb]
\vbox{\includegraphics{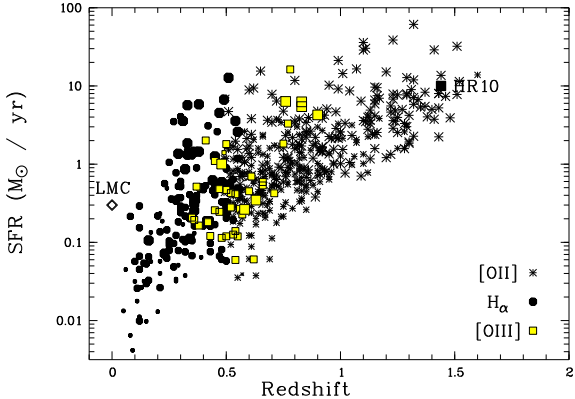}}
\vspace{5.2cm}
                                                                                
\caption{SFR derived from H$\alpha$ (circles), [\ion{O}{3}] (squares) and
[\ion{O}{2}] luminosities (asterisks). Size of the symbol is proportional
to the object absolute broad-band magnitude (in F814W or F775W).
The current H$\alpha$-derived SFR
for the Large Magellanic Cloud (LMC) is indicated by an open diamond
(Kennicutt, Tamblyn \& Congdon 1994).  A filled square illustrates the
[\ion{O}{2}]-derived SFR of the dusty, red starburst galaxy HR10
\citep{Dey:99}.
}

\label{f:SFR_Z}
\end{figure}

We plot the derived star formation rates\footnote{
Note that since our fluxes are not corrected for any absorption effect,
the values we present are a low limit to the true SFRs.}
against redshift in Fig.~\ref{f:SFR_Z}.
At each redshift, our dynamic range in observed SFR
is approximately 1.5 orders of magnitude, and we see the expected bias
of finding higher SFRs at higher redshifts.
The majority of the surveyed
H$\alpha$ emitters are mildly star-forming, local galaxies.  The median
SFR of H$\alpha$-emitters in our survey is $0.2\, {\cal M}_{\odot}\, {\rm
yr}^{-1}$.  After correcting for \nii\ contamination using the average
[\ion{N}{2}]:H$\alpha$ ratio of 0.3 derived by \citet{Gallego:97} from
a local sample of galaxies, this rate is reduced to $\sim 0.1\, {\cal
M}_{\odot}\, {\rm yr}^{-1}$.
This SFR is typical of local dwarf
galaxies:  \eg the median SFRs of the local blue compact galaxies
is about $0.3\, {\cal M}_{\odot}\, {\rm yr}^{-1}$
\citep{Hopkins:02}, and
the H$\alpha$-derived SFR for the nearest irregular galaxies,
LMC and SMC are, respectively,
0.26 and $0.046\, {\cal M}_{\odot}\, {\rm yr}^{-1}$
\citep{Kennicutt:94}.

Using equation (4), we determine that the SFR of [\ion{O}{2}]-emitters
in our survey span nearly 3 orders of magnitude, from approximately a
few times $10^{-2}$ to several times $10^1\, {\cal M}_{\odot}\, {\rm yr}^{-1}$,
with a median SFR of about $1.2\, {\cal M}_{\odot}\, {\rm yr}^{-1}$.
Our survey
is sensitive enough to detect objects with SFRs as low as $1\, {\cal
M}_\odot\, {\rm yr}^{-1}$ up to $z \approx 1.2$.
At higher-redshift the ACS grism survey reaches typical starburst
galaxies with SFRs of $20-60 \, {\cal M}_{\odot}\, {\rm yr}^{-1}$
\citep[\eg][]{Glazebrook:99,Savaglio:04}.
Even a distant analog of the ULIRG population,
such as ERO J164502+4626.4 --- also known as HR10
\citep[${\rm SFR}_{\rm [OII]}
\approx 10\, {\cal M}_{\odot}\, {\rm yr}^{-1}$;][]{Dey:99} ---
can be among our high-redshift sample.


A major concern remains the large uncertainty in SFRs derived from oxygen
fluxes.  It has previously been noted in surveys of the local universe
that the [\ion{O}{2}]:H$\alpha$ ratio correlates with total galaxy
luminosity, such that brighter galaxies have lower [\ion{O}{2}]:H$\alpha$
ratios \citep[e.g.][]{Jansen:01, Tresse:02}.  This correlation is thought to
be related to metallicity.  The variation in the [\ion{O}{3}]:H$\alpha$
ratio is even more dependent on metallicity, as well as on the effective
temperature of the gas and the ionization parameter \citep{Kennicutt:00},
so we proceed with this caution in mind.  We note, however, that the
[\ion{O}{2}] and [\ion{O}{3}] measurements of the SFR do not show a large
discontinuity with the H$\alpha$ measurements at the transitional region
of $z\approx0.5$ (Fig.~\ref{f:SFR_Z}); apparently, the SFR determinations
from oxygen lines are not completely wrong.

\begin{figure}[tb]
\vbox{\includegraphics{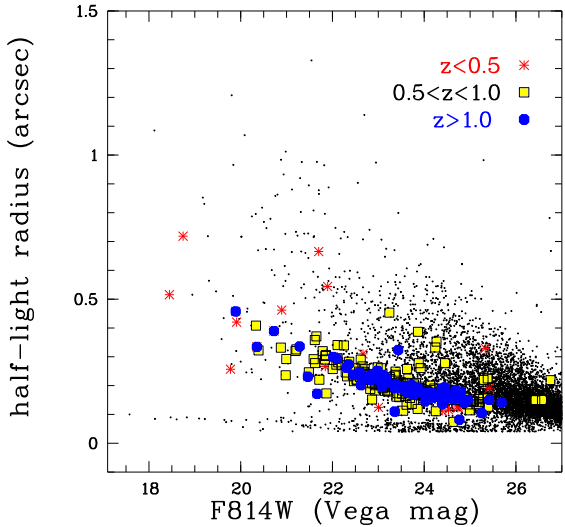}}
\vspace{6.9cm}

\caption{Half-light radius versus total magnitude for three ELG redshift
bins (large symbols, as indicated).  For comparison, all sources in our
direct images are plotted as small dots.  The concentration of the ELGs
among the objects of small angular size objects is evident.}

\label{f:F814vsHLR}
\end{figure}

A rough estimate of the number density of star-forming galaxies
can be made using
galaxies detected at $0.3 \leq z \leq 1.3$, such that their H$\alpha$
and [\ion{O}{2}] lines are in efficient regions of the sensitivity
curve.  The total angular area of the eleven selected fields is 121
arcmin$^2$, corresponding to a co-moving volume of $1.13 \times
10^5\, h^{-3}_{70}$~Mpc$^{3}$.  We detected 506 ELGs in this
redshift interval, giving a co-moving number density of $4.5 \times
10^{-3}\, h^{-3}_{70}$~Mpc$^{-3}$.  The co-moving number density of
[\ion{O}{2}]-emitters in our survey at $0.5 < z < 1.3$ is $\sim 3.8 \times
10^{-3}\, h^{-3}_{70}$~Mpc$^{-3}$.  This density is about 7 times higher
than that detected by the STIS parallels \citep{Teplitz:03} at $0.5 <
z < 1.2$.  The NICMOS parallel survey \citep{McCarthy:99} detected
33 H$\alpha$ emitters at $0.7 < z < 1.9$ in the co-moving volume of
$0.78 \times 10^5\, h^{-3}_{70}$~Mpc$^{3}$.  Thus the co-moving number
density of ELGs averaged over their volume is $\sim 0.4 \times 10^{-3}\,
h^{-3}_{70}$~Mpc$^{-3}$.
As \citet{Yan:99} and \citet{Teplitz:03}
point out, their surveys probed starburst galaxies from the upper
end of the luminosity function.  Our data set is able to detect much
fainter emission lines and measure the co-moving star formation density
at $z<1.6$ to better completeness levels.  On the other hand,
the redshifts measurements based on the ACS/WFC slitless spectra are subject
to larger uncertainties due to the lower spectral resolution of ACS as
compared to STIS and NICMOS\footnote{
The deep, targeted NICMOS survey of the 4.4 arcmin$^2$ Groth-Westphal strip
\citep{Hopkins:00} detected 37 H$\alpha$ emitters
at $0.7<z<1.8$, corresponding to the co-moving number density of
$\sim 7.3  \times 10^{-3}\, h^{-3}_{70}$~Mpc$^{-3}$.
We also expect that the GRAPES ACS
grism survey of the Hubble Ultra Deep Field \citep{Pirzkal:04}, with
total grism time $\sim28$ hours, will reveal a much larger number
density of emission-line galaxies within the same redshift range
as our survey.}.


\begin{figure}[tb]
\vbox{\includegraphics{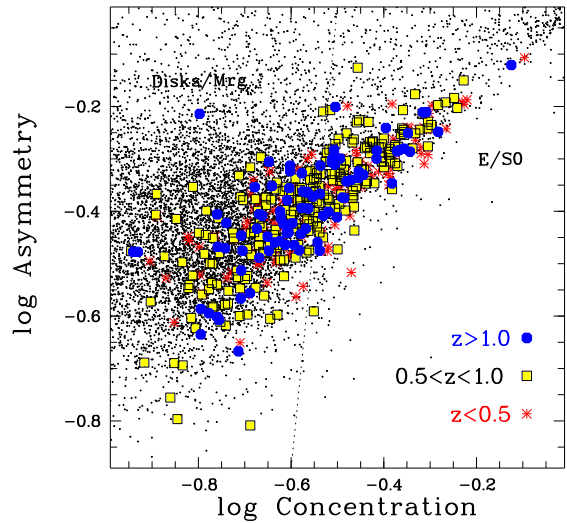}}
\vspace{6.9cm}

\caption{The distribution of ELGs on the central concentration $-$
asymmetry ($C - A$) plane,  as measured in F775W/F814W images.
Large symbols represent ELGs; dots show all sources detected in the
direct images.  The dotted line is a rough border between disk and
bulge dominated galaxies, defined from the visual classification of
well-resolved, bright galaxies.
}

\label{f:lgClgA}
\end{figure}

\subsection {The morphology of the ELGs}

The high spatial resolution of the ACS/WFC images enable a detailed
study of the morphological properties of ELGs.  We defer a detailed
study to a future paper and present here plots of basic morphological
parameters for the ELG sample, such as half-light radius, concentration,
and asymmetry.  We analyze F775W/F814W images (observed $I$-band),
which correspond to rest-frame $B$-band for the median redshift of
our ELG sample, $\langle z \rangle \sim 0.7$.  The morphology at
this rest-frame wavelength is dominated by star-formation regions.
Fig.~\ref{f:F814vsHLR} plots half-light radius against total magnitude
for all objects detected in our imaging, with identified ELGs indicated.
Because of band-shifting effects and surface brightness dimming, plot
symbols are keyed to redshift.  As expected, the survey is biased towards
obtaining redshifts for compact galaxies.

The morphological classifications of the ELGs galaxies were performed
on the basis of both the visual inspection and automatic classification
using central concentration ($C$) and rotational asymmetry ($A$)
indexes \citep{Abraham:94, Abraham:96}.  As shown by \citeauthor{Abraham:96},
these parameters are remarkably robust to image degradation resulting from
increased line-of-sight distance.  Visual classification, however, is more
sensitive to peculiar and merging galaxies.  The comparison of visual
with quantitative morphological classifications show good agreement in
$\sim 70\%$ of galaxies brighter than F814W$= 22$.  For fainter galaxies,
it is not trivial to distinguish E from S0 galaxies, and Sb or Sc spirals
from Sd and Irr galaxies. Therefore, based on the visual calibration of
well-resolved, bright galaxies, we distinguish here two classes of objects:
disk-dominated and bulge-dominated galaxies.

The distribution of the entire imaging data set on the
$C - A$ plane is shown in Fig.~\ref{f:lgClgA},
revealing
the variety of their morphological types. The fraction
of disk versus bulge dominated objects in our sample is
almost equal.
The $C - A$ plot is also
characterized by a correlated distribution of the ELGs, which can be well
approximated by $\log A = (0.53\pm0.03) \cdot \log C - (0.08\pm0.02)$.
The small fraction of ELGs among highly asymmetric galaxies might be
explained by our bias towards galaxies of small angular size.  The less
concentrated disk-like systems show a larger spread in asymmetry, with
many of them resembling interacting, merging, and peculiar systems.

%

\subsection {ACS simulations and completeness tests}

\begin{figure*}[tb]
\vbox{\includegraphics{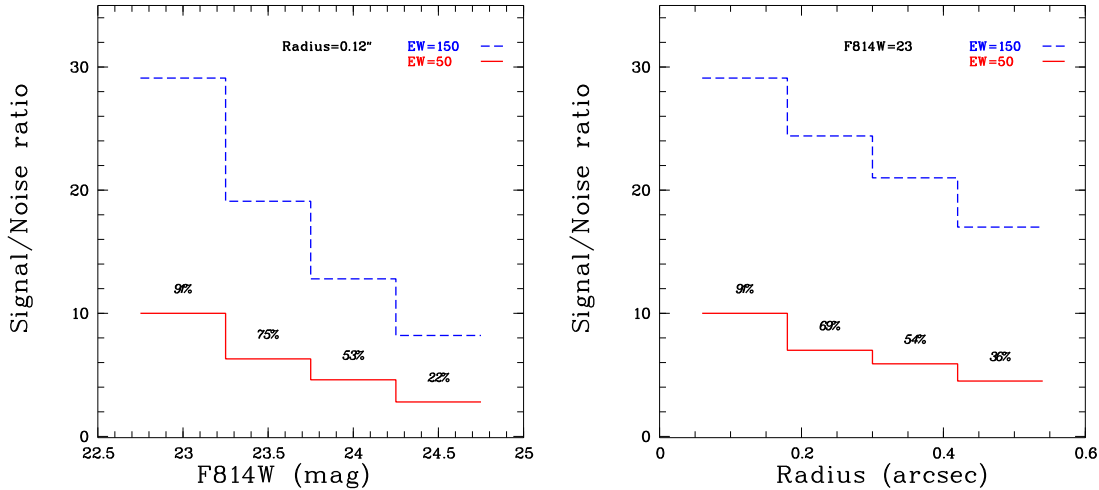}}
\vspace{6.8cm}

\caption{Simulated emission line signal-to-noise ratio (S/N), as a
function of galaxy magnitude (left) and radius (right).
The solid line is the mean S/N of an [\ion{O}{2}] emission line
with an equivalent width (EW) of 50~\AA. The dashed line is same for an
[\ion{O}{3}] emission-line with EW$=150$~\AA\ , corresponding to the SB1
spectral template
\citet{Kinney:96} redshifted to $z=0.7$.  The fraction of
[\ion{O}{2}]-sources detected at $>3\sigma$ (in
percent) are shown in italics.}

\label{f:compl}
\end{figure*}

There are several aspects of the data which negatively impact our
ability to identify genuine spectral features. Since we are observing
in slitless mode, each object produces zero, first and second order
spectra. The first order spectra contain the most useful data; the second
order is profitable for bright objects; zero, negative and higher ($>
2^{\rm nd}$) orders are sources of confusion.  The zero-order images
can can be mistaken for emission features, particularly when they fall
on the first order continua of other objects.  The zero-order images
are slightly dispersed ($\sim 650$~\AA\ pix$^{-1}$) and often appear
bimodal for point sources.  The displacement between the zero order
images and the center of the corresponding direct image (near the
start of the first order spectrum) is $\sim 5\farcs5$, or $\sim 110$
pix, and generally can be identified by matching them with either the
first order spectra or with an object in the direct image.  There is
a small portion of the detector, however, for which zero-order images
can appear {\em without} either first order spectra or images in the
direct frame. In addition to the confusion caused by spectral orders,
there are artifacts associated with defective pixels and cosmic rays.

We performed completeness tests using artificial object trials with the
{\tt SLIM} program \citep{Pirzkal:01}, using a set of template starburst
(SB1) galaxy spectra from \citet{Kinney:96}.  The original {\tt SLIM}
configuration parameters were adjusted to correspond to the current best-known
ACS/WFC/G800L grism spectral trace and dispersion descriptions.  A total
of $\sim 600$ pairs of the synthetic direct images and $z = 0.7$ SB1
template spectra were generated with {\tt SLIM} and randomly added to
our F814W and G800L data for the J~13:58$+$62:39 field.  The synthetic
images used Gaussian brightness profiles with radii $0\farcs1 \leq r \leq
0\farcs5$ and magnitudes in the range $23.0 \simlt$ F814W $\simlt 24.5$.
Object spectra were extracted using the same method we used for the real
objects and the emission-line parameters were measured and compared with
input ones.

Objects were considered recovered if they were found in both direct
and grism images, the emission lines were detected at the $>3\sigma$
level, and the measured equivalent widths did not exceed the initial,
injected values by 20~\AA.  Fig.~\ref{f:compl} shows the outcome of these
completeness tests.  These plots reveal strong selection effects are
present in our sample of ELGs.  More detailed tests, necessary for the
calculation of the luminosity function and measurements of co-moving
star formation rate, will be discussed in Paper~III.

The quality of the flux and wavelength calibration have also been tested
on data taken at different epochs and located at different parts of
the ACS/WFC detector:  we find good agreement.  The rms accuracy of
redshifts derived from repeat observations is $\sim 0.02$.

\subsection {Notes on Individual Fields}

\subsubsection {Field J~13:58+62:39}

This field has the largest total grism exposure time of all the survey
fields, and reveals many interesting objects.  We identify more than 40
ELGs, including a concentration of galaxies at $z \sim 0.3$.

\subsubsection {Field J~12:19+06:49}

This field was parallel to a program aiming at a Virgo cluster galaxy,
located in the outer regions of the cluster.  We detect a small, low surface
brightness spheroidal galaxy at 12:19:10.96, +06:47:54.0 (J2000), which
has a bright, compact source at its optical center with blue-excess
continuum. However, it is unclear if the central object is associated with
the galaxy. Based
on the galaxy patchy faint background traced out to galactocentric radii
of $\approx10\arcsec$ in both $F775W$ and $F850LP$-band images,
the galaxy appears to be an isolated spheroidal galaxy, likely at the
distance of the Virgo cluster. At the same time, we could not rule out the
possibility that this is a Galactic source (\eg planetary nebula).
No record of this object exists in NED or SIMBAD.

\subsubsection {Field J~10:03+29:06}

This field shows a surprisingly large number of ELGs, despite its short total
grism exposure time. Eleven emission line galaxies and several candidates
high-redshift \lya\ emitters were observed with Keck/LRIS, as discussed
in \S\ref{s:lris}; however, conditions for this ground-based spectroscopy
were insufficient for obtaining useful data on any but the brightest galaxies.

\subsubsection {Field J~08:06+06:43}

The ACS data for this field is part of the GO program 9405
(P.I. Andrew Fruchter), which targeted the host galaxy of
supernova SN~2002LT, associated with gamma-ray burst GRB021211
\citep{Crew:03}. The ACS grism spectrum confirms the presence of an emission
line at $\sim7450$~\AA\, at the position of the probable host galaxy
(J~08:08:59.828, +06:43:37.52), as previously detected in a VLT/FORS2
spectrum by \citet{Vreeswijk:03}. This line is likely \oii\,
at z=1.006.
The deep multi-epoch slitless and direct data allow us to detect 88 ELGs
in this field.

\section{Discussion}

We present basic data derived from the ACS Grism Parallel Survey.  The
G800L grism on ACS provides a unique opportunity to survey large volumes
of the universe for faint emission lines and at high angular resolution.
The small pixel scale of the ACS images and our custom software for
deblending object spectra provides us with the unique opportunity to identify
faint star-forming regions across vast cosmic epochs.  The large dataset
afforded by our ``random'' parallel observations allows us to collect
data from much larger area than would be possible in a single GO program.

The ACS Grism Parallel Survey complements previous and ongoing surveys
of [\ion{O}{2}], [\ion{O}{3}] and H$\alpha$ ELGs.  Our faint flux limits
allow us to probe deeper into the ELG luminosity function.  In this paper
we present our methodology for data analysis and the first results from
our survey.
Our initial survey of 121 square arcminutes detects 601 emission-line
galaxies at redshifts $z \leq 1.6$.
The line luminosities, equivalent
widths and continuum magnitudes suggest that we are seeing galaxies with
a broad range of star formation rates, from quiescently star-forming
galaxies at low redshifts to bright starburst galaxies at $z>1$.

The survey is biased towards compact objects with strong emission lines.
Such galaxies tend to be either starbursts and/or AGNs.
Follow-up high-resolution spectra are necessary in order to classify
each ELG by its activity. To date we have obtained follow-up spectra
for 11 candidates, all of them found to be starforming ELGs with seven
being starbursts (see \S\ref{s:lris}).
While this subsample is small, we infer that the
fraction of AGN in our sample is small and that
the inferred emission-line luminosities can be used to
estimate the co-moving star formation density at $z \leq 1.6$ at a
high confidence level.

\acknowledgements
The authors also wish to recognize and acknowledge the very
significant cultural role and reverence that the summit of Mauna Kea
has always had within the indigenous Hawaiian community; we are most
fortunate to have the opportunity to conduct observations from this
mountain.  ID acknowledges the support from NASA HST grant GO-9468.
The work of DS was carried out at Jet Propulsion Laboratory,
California Institute of Technology, under a contract with NASA.


\begin{deluxetable}{crcrrrrrrrrrrrrrr}
\tabletypesize{\footnotesize}
\rotate
\tablewidth{0pt}
\tablecaption{The ACS Grism Survey Fields
  \label{t:Fields}}
\tablehead{
	       & \colhead{ } & \multicolumn{2}{c}{G800L} & \multicolumn{3}{c}{F814W}& \multicolumn{3}{c}{F775W} & \multicolumn{3}{c}{F850LP}& \multicolumn{3}{c}{F606W} & \colhead{     }\\
		\colhead{Field} & \colhead{b\tablenotemark{a}}
	       & \multicolumn{1}{c}{\#Exp\tablenotemark{b}} & \multicolumn{1}{c}{Time\tablenotemark{c}}
	       & \multicolumn{1}{c}{\#Exp\tablenotemark{b}} & \multicolumn{1}{c}{Time\tablenotemark{c}}&\colhead{$m(\sigma)$\tablenotemark{d}}
	       & \multicolumn{1}{c}{\#Exp\tablenotemark{b}} & \multicolumn{1}{c}{Time\tablenotemark{c}}&\colhead{$m(\sigma)$\tablenotemark{d}}
	       & \multicolumn{1}{c}{\#Exp\tablenotemark{b}} & \multicolumn{1}{c}{Time\tablenotemark{c}}&\colhead{$m(\sigma)$\tablenotemark{d}}
	       & \multicolumn{1}{c}{\#Exp\tablenotemark{b}} & \multicolumn{1}{c}{Time\tablenotemark{c}}&\colhead{$m(\sigma)$\tablenotemark{d}}
	       & \colhead{Prog.}
}
\startdata
 J\,$01\,22-28\,24$ &-83& 25&12825 &   &     &    &  5& 2437&29.6&  5& 2530&28.8&  &    &    & 9482 \\
 J\,$01\,30-16\,04$ &-76& 41&18525 &   &     &    & 11& 5768&30.3&  8& 3700&29.3&  &    &    & 9482 \\
 J\,$02\,27-40\,55$ &-66& 26&12218 & 13& 6195&30.3&   &     &    &   &     &    &  &    &    & 9468 \\
 J\,$07\,26+69\,15$ & 28& 22&15526 &   &     &    &  6& 3200&29.8&  6& 3060&29.1&  &    &    & 9482 \\
 J\,$08\,08+06\,43$ & 20& 22&22390 & 24&11680&30.8&   &     &    &   &     &    &39&15580&30.9& 9405 \\
 J\,$10\,03+29\,06$ & 53& 27&13481 & 17& 8750&30.5&   &     &    &   &     &    &  &    &    & 9468 \\
 J\,$11\,29-14\,39$ & 44& 26&17398 &   &     &    &  2& 1000&29.2& 19&12911&30.0&  &    &    & 9482 \\
 J\,$12\,19+06\,49$ & 68& 25&16800 &   &     &    & 20&10938&30.5& 15& 9798&29.7&  &    &    & 9482 \\
 J\,$13\,39+00\,08$ & 61& 27&13678 &   &     &    &  8& 4460&29.9&  8& 3760&29.1&  &    &    & 9482 \\
 J\,$13\,58+62\,39$ & 53& 42&24319 & 18&10226&30.7&  4&     &    &  2& 1000&29.0&  &    &    & 9468 \\
 J\,$15\,42-10\,46$ & 34& 43&24024 &   &     &    & 21&11586&30.1& 20&10669&29.8&  &    &    & 9482 \\
\enddata
\tablenotetext{a}{Galactic latitude, in degrees.}
\tablenotetext{b}{Number of images and spectra.}
\tablenotetext{c}{The total exposure time, in seconds.}
\tablenotetext{d}{
The $1\sigma$ sky $rms$ magnitude (Vega system), measured in the
aperture of $1\times FWHM$ diameter (0.1~arcsec).
In each filter, the number represents the average value for a whole area
of the combined frame.}
\end{deluxetable}

\begin{deluxetable}{rrrrrrrrr}
\tabletypesize{\tiny}
\tablewidth{0pt}
\tablecaption{Imaging Properties of Emission-Line Objects
  \label{t:Phot}}
\tablehead{
	     & \colhead{RA}        & \colhead{DEC}       & \colhead{$F814W$} & \colhead{a}        & \colhead{b}        & \colhead{$\theta$} & \colhead{     }& \colhead{     } \\
	{Id}  & \colhead{(h\,m\,s)} & \colhead{(d\,m\,s)} & \colhead{(mag\tablenotemark{a})} & \colhead{(pix)} & \colhead{(pix)} & \colhead{(deg)} & \colhead{Conc.}& \colhead{Asym.}
}
\startdata
410 &13:58:46.190  &+62:37:59.65 &23.54$\pm$0.01 &2.3 &2.1 &-60.8 &0.484 &0.615 \\
411 &13:58:47.083  &+62:38:05.91 &23.82$\pm$0.05 &8.1 &3.2 & 45.4 &0.302 &0.341 \\
412 &13:58:49.389  &+62:39:05.53 &22.84$\pm$0.01 &3.6 &2.4 &-88.4 &0.430 &0.521 \\
413 &13:58:51.968  &+62:40:49.81 &24.51$\pm$0.03 &3.0 &1.6 & 83.5 &0.249 &0.430 \\
414 &13:58:52.446  &+62:40:02.95 &25.70$\pm$0.06 &1.5 &1.3 &-41.5 &0.421 &0.492 \\
415 &13:58:53.549  &+62:40:27.43 &24.45$\pm$0.02 &2.3 &1.7 & 69.3 &0.398 &0.548 \\
416 &13:58:54.556  &+62:41:02.73 &23.15$\pm$0.01 &2.0 &1.8 &-54.1 &0.592 &0.708 \\
417 &13:58:54.327  &+62:39:50.81 &23.63$\pm$0.01 &3.0 &2.3 &-83.0 &0.321 &0.466 \\
418 &13:58:59.669  &+62:37:51.34 &23.00$\pm$0.01 &3.5 &2.5 &-75.5 &0.492 &0.570 \\
419 &13:59:00.595  &+62:40:39.99 &24.61$\pm$0.03 &2.8 &1.6 & -4.3 &0.411 &0.462 \\
420 &13:58:57.771  &+62:38:09.23 &22.75$\pm$0.01 &4.7 &2.7 & 55.0 &0.394 &0.512 \\
421 &13:59:00.284  &+62:40:14.23 &23.46$\pm$0.01 &3.2 &2.0 &-63.2 &0.402 &0.574 \\
422 &13:58:57.438  &+62:38:21.04 &24.72$\pm$0.02 &1.8 &1.6 &-20.3 &0.341 &0.539 \\
423 &13:58:57.490  &+62:38:38.09 &22.13$\pm$0.01 &7.2 &2.8 &-85.9 &0.350 &0.494 \\
424 &13:58:57.189  &+62:39:08.42 &23.90$\pm$0.02 &3.2 &2.1 &-41.4 &0.289 &0.445 \\
\enddata
\tablenotetext{a}{SYNPHOT standard Vega magnitude}
\tablecomments{The complete version of this table is in the electronic
edition of the Journal.  The printed edition contains only a sample.}
\end{deluxetable}

\begin{deluxetable}{rrrrrrrrrrrrrrcc}
\rotate
\tabletypesize{\tiny}
\tablewidth{0pt}
\tablecaption{Spectroscopic Properties of Emission-Line Objects
  \label{t:Spec}}
\tablehead{
	      &             & \multicolumn{4}{c}{H$_{\alpha}$} & \multicolumn{4}{c}{H$_{\beta}+$[OIII]} & \multicolumn{4}{c}{[OII]} &  &                \\
\colhead{Id} &\colhead{$z$}& \colhead{Flux} & \colhead{EW} & \colhead{S/N} & \colhead{cont.} & \colhead{Flux} &  \colhead{EW} & \colhead{S/N} & \colhead{cont.} & \colhead{Flux} & \colhead{EW} & \colhead{S/N} & \colhead{cont.} & \colhead{R\tablenotemark{a}}& \colhead{Comment.\tablenotemark{b}}
}
\startdata
410 & 1.09&                &                 &      &    &                 &               &      &    &  69.9$\pm$ 9.9 & 251.8$\pm$35.6&   7.1& 0.3 &b&            \\
411 & 0.31&  44.9$\pm$ 5.8 & 141.6$\pm$18.3  &   7.7& 0.3&  32.7$\pm$ 8.3  &  61.3$\pm$15.6&   3.9& 0.5&                &               &      &     &b&HII?        \\
412 & 1.15&                &                 &      &    &                 &               &      &    &  92.4$\pm$30.8 &  92.1$\pm$30.7&   3.0& 0.8 &c&            \\
413 & 0.69&                &                 &      &    &  41.7$\pm$ 5.0  & 338.1$\pm$40.6&   8.3& 0.1&   8.7$\pm$ 2.9 &  48.6$\pm$16.2&   3.0& 0.1 &b&            \\
414 & 0.71&                &                 &      &    &  31.4$\pm$ 1.2  & 510.8$\pm$20.1&  25.4& 0.0&                &               &      &     &a&BCG         \\
415 & 0.83&                &                 &      &    &  60.7$\pm$ 3.5  & 363.3$\pm$21.1&  17.2& 0.1&   7.4$\pm$ 2.5 &  34.7$\pm$11.6&   3.0& 0.3 &c&S           \\
416 & 0.62&                &                 &      &    &  64.0$\pm$21.3  &  93.6$\pm$31.2&   3.0& 0.7&  91.3$\pm$30.4 & 113.6$\pm$37.9&   3.0& 0.8 &c&BCG?        \\
417 & 0.32&  29.0$\pm$ 6.1 &  88.4$\pm$18.6  &   4.8& 0.3&  55.5$\pm$14.1  & 109.9$\pm$28.0&   3.9& 0.5&                &               &      &     &a&            \\
418 & 0.37&                &                 &      &    & 184.9$\pm$12.4  & 291.5$\pm$19.5&  15.0& 0.6&                &               &      &     &a&            \\
419 & 0.01& 111.6$\pm$ 2.8 & 956.7$\pm$23.6  &  40.5& 0.1&                 &               &      &    &                &               &      &     &b&BCG?        \\
420 & 0.29& 177.4$\pm$59.1 & 253.9$\pm$84.6  &   3.0& 0.6& 130.1$\pm$43.4  & 132.0$\pm$44.0&   3.0& 1.0&                &               &      &     &a&            \\
421 & 1.32&                &                 &      &    &                 &               &      &    &  48.0$\pm$16.0 & 141.6$\pm$47.2&   3.0& 0.3 &c&            \\
422 & 0.69&                &                 &      &    & 115.7$\pm$ 3.4  &1606.3$\pm$47.6&  33.8& 0.1&  11.4$\pm$ 1.6 & 132.4$\pm$18.8&   7.0& 0.1 &b&BCG         \\
423 & 0.85&                &                 &      &    & 293.8$\pm$21.7  & 359.2$\pm$26.6&  13.5& 0.8&  44.9$\pm$15.0 &  29.2$\pm$ 9.7&   3.0& 1.5 &b&            \\
424 & 0.98&                &                 &      &    &                 &               &      &    &  40.7$\pm$ 5.4 & 158.4$\pm$20.9&   7.6& 0.3 &b&BCG         \\
\enddata
\tablecomments{All observed fluxes are in unites of $10^{-18}$~ergs cm$^{-2}$ s$^{-1}$, and equivalent widths are in \AA.
Fluxes and equivalenth widths are not corrected to the rest frame and absorption. }
\tablenotetext{a}{Reliability of line identification: `a'='Good', `b'=`OK', `c'=`Uncertain'}
\tablenotetext{b}{Symbols are as follows:
`BCG'=`Blue Compact Galaxy', `S'=`Spiral', `Ir'=`Irregular',
`HII'=`HII region', `\#c'=`\# components'.}
\tablecomments{The complete version of this table is in the electronic
edition of the Journal.  The printed edition contains only a sample.}
\end{deluxetable}


\end{document}